\documentclass[twocolumn]{svjour3}          % twocolumn
\smartqed  % flush right qed marks, e.g. at end of proof
\usepackage{amsmath, amssymb}
\usepackage{graphicx} % Include figure files
\usepackage{dcolumn} % Align table columns on decimal point

\newcommand{\APMLab}{
    State Key Laboratory of Magnetic Resonance and Atomic and Molecular Physics
    Innovation Academy for Precision Measurement Science and Technology, 
    Chinese Academy of Sciences, 
    Wuhan 430071}
\newcommand{\APM}{
    Key Laboratory of Atomic Frequency Standards,
    Innovation Academy for Precision Measurement Science and Technology, 
    Chinese Academy of Sciences, 
    Wuhan 430071}
\newcommand{\UCAS}{
    University of the Chinese Academy of Sciences, 
    Beijing 100049}
\newcommand{\WIQT}{
    Wuhan Institute of Quantum Technology, 
    Wuhan 430206}

\newcommand{\braket}[1]{ \langle #1 \rangle} % braket

\newcommand{\Al}{{}^{27}\textrm{Al}^{+}}
\newcommand{\Ca}{{}^{40}\textrm{Ca}^{+}}
\newcommand{\Mg}{{}^{25}\textrm{Mg}^{+}}
\newcommand{\Be}{{}^{9}\textrm{Be}^{+}}
\newcommand{\s}{{}^1\textrm{S}_0}
\newcommand{\p}{{}^3\textrm{P}_0}
\newcommand{\sCa}{{}^2\textrm{S}_{1/2}}
\newcommand{\pCa}{{}^2\textrm{P}_{1/2}} 

\newcommand{\dCa}{{}^2\textrm{D}_{5/2}}
\newcommand{\ee}{10^{-18}}

\newcommand{\sZeeman}{8617.0}
\newcommand{\sSM}{6.9} 
\newcommand{\sEMM}{3.9}
\newcommand{\sBBR}{3.2}
\newcommand{\sStark}{1.1}
\newcommand{\sAOM}{0.0}
\newcommand{\sBGC}{0.0}
\newcommand{\sDoppler}{0.0}
\newcommand{\sTotal}{8632.2}

\newcommand{\sClock}{0.0} 

\newcommand{\uZeeman}{6.4} 
\newcommand{\uSM}{3.1}  
\newcommand{\uEMM}{2.9}
\newcommand{\uBBR}{0.5}
\newcommand{\uStark}{1.1} 
\newcommand{\uAOM}{0.3} 
\newcommand{\uBGC}{0.2}
\newcommand{\uDoppler}{0.6}
\newcommand{\uTotal}{7.9}

\newcommand{\uClock}{0.1}

\begin{document}

\title{ 
    Evaluation of the systematic shifts of a $\Ca-\Al$ optical clock
}

\author{Kaifeng Cui$^{1,2,\ddag}$ \and
        Sijia Chao$^{1,2,*}$ \and
        Chenglong Sun$^{1,2,3}$ \and
        Shaomao Wang$^{1,2,3}$ \and
        Ping Zhang$^{1,2,3}$ \and
        Yuanfei Wei$^{1,2,3}$ \and
        Jinbo Yuan$^{1,2}$ \and 
        Jian Cao$^{1,2}$ \and
        Hualin Shu$^{1,2}$ \and
        Xueren Huang$^{1,2, 4, \dagger}$
}
\authorrunning{K. Cui et al.}

\institute{
    $^1$\APM.  \\
    $^2$\APMLab. \\ 
    $^3$\UCAS. \\
    $^4$ \WIQT\\
    ${^*}$ These authors contribute equally. \\
    $^\ddag$\email{cuikaifeng@apm.ac.cn}
    $^\dagger$\email{hxueren@apm.ac.cn}.    
}
\date{\today}

\maketitle

\begin{abstract}

Quantum-logic-based $\Al$ optical clock has been demonstrated in several schemes as there are different choices of the auxiliary ion species.
In this paper, we present the first detailed evaluation of the systematic shift
    and the total uncertainty of an $\Al$ optical clock
    sympathetically cooled by a $\Ca$ ion.
The total systematic uncertainty of the $\Ca - \Al$ quantum logic clock 
    has been estimated to be $\uTotal \times \ee$,
    which was mainly limited by the uncertainty of the quadratic Zeeman shift.
By comparing the frequency of two counter-propagating clock beams on the same ion, we measured the frequency stability to be $3.7 \times 10^{-14} /\sqrt{\tau}$.


\end{abstract}


With the rapid development of optical clocks, a new definition of the SI unit of time using the optical clocks has become realistic~\cite{riehle_2015_redefinition,lodewyck_2019_definition}.
Impressed by their high performance, a wide range of applications in the study of fundamental physics%
    ~\cite{safronova_2018_search}
    such as searching for dark matter~\cite{kennedy_2020_precision,wcislo_2018_new} or
    testing general relativity~\cite{chou_2010_optical,takamoto_2020_test}, 
    have been proposed.
Among all the atomic species, single-ion optical clock based on the $\s \leftrightarrow \p$ transition on $\Al$ has long been considered as a good candidate~\cite{Yu_1992}, not only because of its 8 mHz narrow linewidth, but also for its very low sensitivity to the black-body radiation~(BBR).
Recently, research reported the fractional frequency uncertainty of an $\Al$ optical clock reached $9.4\times 10^{-19}$~\cite{brewer_2019_clock},
    two orders of magnitude better than the best Cesium fountain clock to date~\cite{weyers_2018}.

Although optical clocks based on trapped $\Al$ have outstanding performance, they face unique issues in cooling and detection.
Due to the lack of commercially available ultraviolet laser at the wavelength of 167 nm,
$\Al$ single-ion optical clocks depend on another kind of co-trapped ions to provide 
    sympathetic cooling~\cite{kielpinski_2000_sympathetic} 
    and quantum logic readout~\cite{schmidt_2005_spectroscopy}. 
The first $\Al$ clock was operate with $^9$Be$^+$ as auxiliary ion~\cite{rosenband_2008_frequency} while later the $\Mg$/$\Al$ clock reached the lowest uncertainty to date~\cite{brewer_2019_clock}.
Since then, various number of group~\cite{guggemos_2015_sympathetic,hannig_2019,ma_2020} have started new $\Al$ optical clocks including ours at APM~\cite{cui_2018_sympathetic,chao_2019_observation}.

In this paper, we present a detailed evaluation of an $\Al$ single-ion optical clock
    sympathetically cooled by a $\Ca$ ion.
Comparing to the similar $\Al$ clock sympathetically cooled by $\Mg$ or $\Be$ at the Doppler cooling limit~\cite{rosenband_2008_frequency,chou_2010_frequency},
our result shows a much smaller time dilation shift due to secular motion of the ions,
which agrees with the theoretical expectation~\cite{wubbena_2012_sympathetic}.
We describe the clock system  
    and present the evaluation of the total systematic shift and uncertainty 
    of the clock transition.
    
\begin{figure}
    \includegraphics[width=0.95\linewidth]{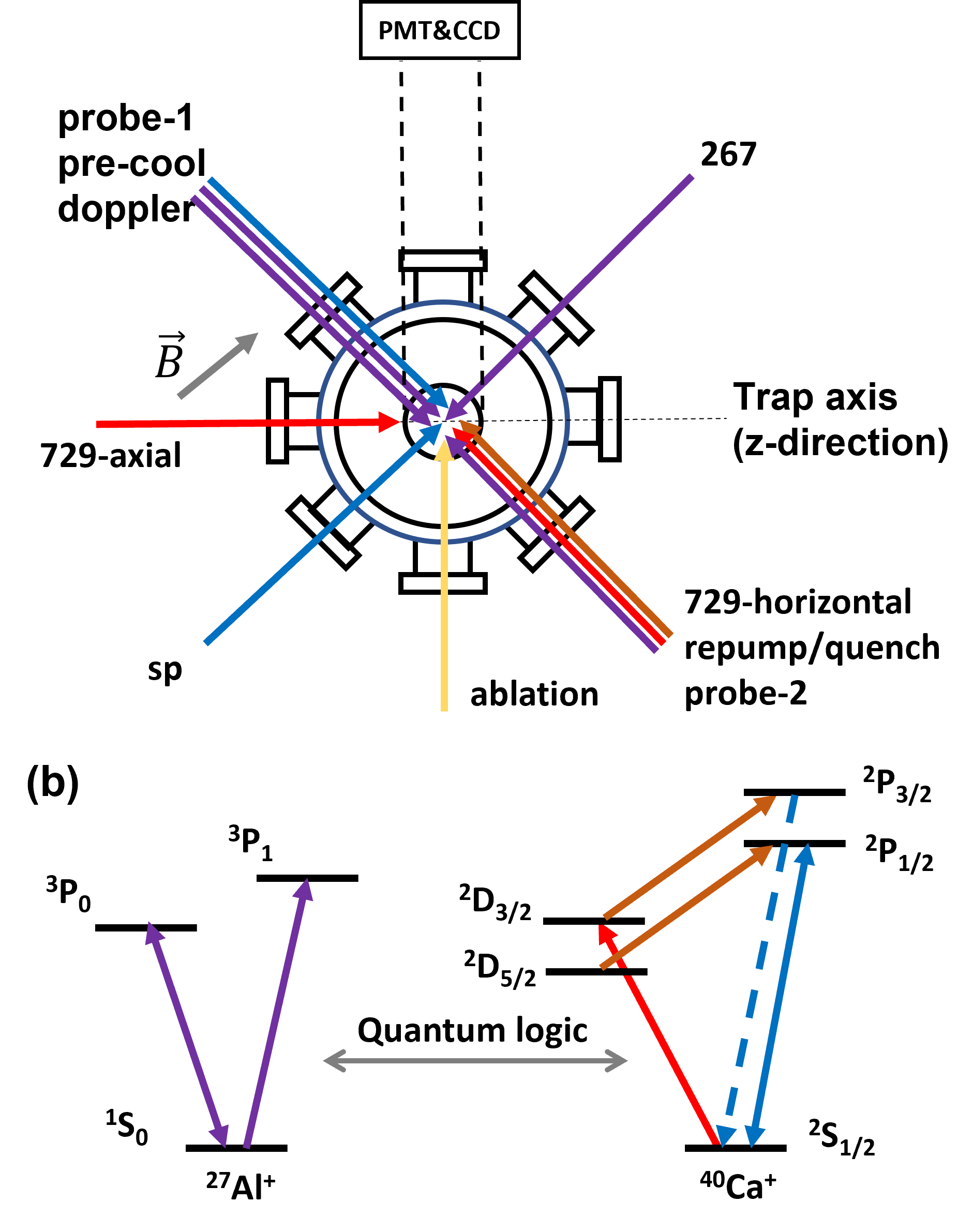}
    \caption{
        (a) Laser beams involved in this experiment.
        A linear Paul trap is placed in the center of the vacuum chamber, with its trap axis along the direction of the 729-axial beam.
        In addition to the 729-axial and 729-horizontal beams,
            there is another 729-vertical beam used for the detection of EMM,
            which is not included in this figure.
        Florescence is collected through an imaging system placed on the vertical direction, sharing the same pathway of the 729-vertical beam.
        SP indicates the state preparation laser for the $\Ca$ ion, while 267 represents a 267 nm laser that prepares the $\Al$ ion and helps with the quantum logic spectroscopy.
        Probe-1 and probe-2 are two almost exactly counter-propagating laser beams that excited the clock transition. When alternately interrogated with each of these beams, possible first-order Doppler shift will be cancelled.
        The stability of the clock is evaluated by comparing the measured frequency of these two beams.
        (b) Energy levels involved in this experiment (Not to scale).
    }
    \label{fig:system}
\end{figure}

We used a linear Paul trap 
    that is similar to the one described previously
    ~\cite{shang_2016_sympathetic,cui_2018_sympathetic,chao_2019_observation} 
    except for that the radio frequency of the trap is increased to 42.9 MHz for a stronger confinement. 
To achieve such a high frequency, the electrodes are made of beryllium copper instead of stainless steel for its lower resistance.
In order to increase the radio frequency (RF) to 42.9 MHz to avoid heating the excess micromotion sideband, the electrodes are made of beryllium copper for its lower resistance than the stainless steel.
The distance between the electrodes to the ion was narrowed to 0.4 mm for stronger confinement.
A pair of end-cap electrodes were placed 8.5 mm apart along the trap axis (z-axis).
A trap frequency of 
    $(\omega_x, \omega_y, \omega_z)$ = (4.1, 4.1, 1.1) MHz was achieved using this trap.

A Nd: YAG laser at 1064 nm with a maximum pulse energy of 15 \textmu J  
    and a pulse duration of 2 ns is used for the ablation loading of 
    both $\Ca$ and $\Al$ ions from two separated metal targets. 
Laser cooling of the $\Ca$ ion is implemented by a 397 nm beam (Doppler beam) 
    that is approximately -10 MHz detuned from resonance.
Another 397 nm pre-cool beam with -120 MHz detuning is co-aligned 
    with the Doppler cooling beam to help the ions recrystallize
    after collision with the background gas, which happens a few times per hour.
A circularly polarized 397 nm laser beam (SP beam) is applied 
    along the magnetic field direction 
    to initialize the $\Ca$ ion to the state $\sCa (m=-1/2)$.
A repump beam at 866 nm from the opposite directions pumps the $\Ca{}$ ion from $^2$D$_{5/2}$ state back to the cooling cycle.
It is co-aligned with 854 nm laser to pump the ion back to the $^2$S$_{1/2}$ state by connecting the $^2$D$_{5/2}$ and $^2$P$_{3/2}$ state.

In the opposite direction, a pair of 267 nm lasers with orthogonal polarization are co-aligned to initialize $\Al$ to $\s (m= \pm 5/2)$ state.
With the help of another 729 nm laser beam along the trap axis (729-axial beam), the internal state of the $\Al$ can be mapped to the shared motional state by the 267 nm laser and measured out using the quantum logic spectroscopy(QLS)~\cite{chao_2019_observation}.
In addition, we employed two more 729 nm beams (729-horizontal and 729-vertical) to detect the excess micromotion in three directions.

The clock transition is probed from two opposite directions.
The first one (probe-1 beam) is co-aligned with the pre-cool beam, and the second one (probe-2 beam) goes together with the repump beam.
These two beams are generated from two separate double passed acousto-optic modulators (AOMs).
The probe-2 beam is aligned well enough to ensure it goes all the way back through the trap and into the AOM of the probe-1 beam, which proves a deviation of the angle smaller than 2$\times 10^{-5}$ rad.

Before each $\Al$ clock interrogation pulse,
    the ions are pre-cooled for 1.5 ms.
A series of 267 nm laser pulses are employed at the same time to pump the $\Al$ ion to the $\s(m_F=\pm 5/2)$ state.
Then a Doppler cooling pulse of 1 ms is applied to cool the ions close to the  Doppler-cooling limit.
The clock interrogation is then applied through either the probe-1 beam or the probe-2 beam.
Both the 397 nm Doppler cooling laser and the 866 nm repump laser are kept on during the clock interrogation pulse
    to ensure that the temperature of the ions remains stable.
A sequence of pulses then 
    maps the $\s$ state of $\Al$ ion to the dark $\dCa(m=-1/2)$ state on $\Ca$
    through their shared motional sidebands~\cite{chao_2019_observation}.
The readout process is repeated 5-20 times using an adaptive Bayesian process~\cite{hume_2007} to reach the lowest measurement error of approximately 0.5\%.

At an interrogation time of 25 ms, we observed a clock transition linewidth of 45 Hz~\cite{chao_2019_observation},
which is larger than the Fourier limit of 32 Hz.
We believe this is limited by the vibration noise in the laser path since the clock laser beam passed two separate optical tables.
For long-term stability, the clock is normally operated at an interrogation time of 10 ms, with a clock transition linewidth of \~80 Hz.

\begin{table}[t] 
    \caption{\label{tab:main}%
    Fractional frequency shifts and uncertainties for the $\Ca$-$\Al$ optical clock in the unit of $10^{-18}$. 
    }
    \begin{tabular}{lcc} % left, center, right
    \hline \noalign{\smallskip}
    \textrm{Effect} &
    \textrm{Shift} &
    \textrm{Uncertainty} \\
    \hline \noalign{\smallskip}
    Quadratic Zeeman         & -$\sZeeman$   & $\uZeeman$ \\
    Secular motion           & -$\sSM$       & $\uSM$ \\
    Excess micromotion       & -$\sEMM$      & $\uEMM$ \\

    Blackbody radiation      & -$\sBBR$      & $\uBBR$ \\
    
    Laser Stark              & -$\sStark$    & $\uStark$ \\
    AOM freq. error          & $\sAOM$       & $\uAOM$ \\
    
    First-order Doppler      & $\sDoppler$   & $\uDoppler$  \\    
    Background-gas collision & $\sBGC$       & $\uBGC$ \\

    Total                    & -$\sTotal$    & $\uTotal$ \\
    \hline \noalign{\smallskip}
    \end{tabular}
\end{table}

The systemic shift is listed in Table~1.
The largest shift arises from the influence of the external magnetic field.
Considering the first-order and second-order terms in the magnetic field, 
    the resonance frequencies of the clock transition will be shifted to ~\cite{brewer_2019_measurements}:
\begin{equation}
    \nu = \nu_0 + C_1 \braket{B} + C_2 \braket{B^2},
\end{equation}
where $\nu_0$ is the unperturbed resonance frequency and 
    $C_1, C_2$ are the coefficient quantifying the linear 
    and the quadratic Zeeman shift, respectively. 
The first order term $\braket{B}$ is compensated 
    by the interleaved locking of two transitions: $\s (m_F=\pm5/2) \leftrightarrow \p (m_F=\pm5/2)$, 
while the second order term, $\braket{B^2} = \braket{B_{DC}}^2 + \braket{B_{AC}^2}$ contributes the largest frequency shift.
The dominant AC components of the magnetic field is the 50 Hz and harmonics $B_{LF}$ and the trap RF induced oscillating magnetic field $B_{RF}$. 
The limited linewidth of the $\s (m_F=+5/2) \leftrightarrow \p (m_F=+5/2)$ transition of the $\Ca$ ion due to the magnetic field noise decoherence is used to infer the lower frequency AC components $B_{LF} = 1.795 \times 10^{-7}$ T.

The oscillating magnetic field $B_{RF}$ modulates the $\sCa \leftrightarrow \dCa$ transition of the $\Ca$ ion with a modulation index $\beta_m = (g_D m_D - g_S m_S)\mu_B B_{RF}/\hbar \Omega_{RF}$~\cite{gan_2018}, where $g_S, g_D, m_S, m_D$ are the g-factors and magnetic quantum number of $\sCa$ and $\dCa$ states, respectively, $\mu_B$ is the Bohr magneton and $\hbar$ is Planck’s constant. However, trap RF induced micromotion, also modulates this transition with index $\beta_{0}$. The measured Rabi rate ratio $\eta$ of the carrier and sideband of the transition contains information of both modulation $\eta = (\beta_{0} + \beta_{m})/2$.
The contribution from the micromotion can be removed by using a pair of transitions with different magnetic quantum numbers.
In our experiment, the micromotion is first minimized and evaluated as described in the following section. Then the 729-axial beam is employed to measure the Rabi rate ratio of $\sCa(m_S = -1/2) \leftrightarrow \dCa(m_D=-1/2)$ transition and $\sCa(m_S = -1/2) \leftrightarrow \dCa(m_D=-5/2)$ gives $\beta_m=2.07\times10^{-3}$, corresponding to $B_{RF}=3.67\times10^{-6}$ T after taking the projection angle of 729-axial beam into account.

As suggested in Ref.~\cite{gan_2018}, this method may ignore the oscillating magnetic field  orthogonal to the quantization axis, results in an underestimate of the total $B_{RF}$ by a factor of 2. Since our current set up does not allow a more precise measurement, we believe taking $B_{RF} = 7.34\times10^{-6}$ T provides an upper bound of $B_{RF}$. With quadratic Zeeman coefficient $C_2$ = $-7.1944(24) \times 10^7  \textrm{ Hz/T}^2$~\cite{brewer_2019_measurements}, 
the fractional frequency shift due to the AC component of the magnetic field is estimated to be below 1.26 $\times 10 ^{-18}$.
Since $B_{DC} = 3.6640(12) \times 10^{-4}$ T can be measured with high precision, the total fractional frequency shift due to the quadratic Zeeman effect is 
    $-(\sZeeman \pm \uZeeman) \times \ee$, limited by the uncertainty of $C_2$.
It is noticeable that reducing the magnetic field strength $\braket{B^2}$ will lead to significant improvement of the clock performance in the future. Although a much smaller magnetic field increases the coupling to other unwanted Zeeman components in the $\sCa \leftrightarrow \dCa$ transition of $\Ca$, results in a larger measurement error in the QLS process.

\begin{table*}
    \caption{\label{tab:sm_result}
    One of the measurement results of the average motional quantum number $\overline{n}_m$ that is used to calculate the corresponding time-dilation shift.
    $z$ stands for the amplitude of the motion at the ground state.
    $\overline{n}_c$ is the calculated average motional quantum number at Doppler cooling limit.
    TDS/quantum is the calculated time-dilation shift per quantum number 
        that has included the contribution of intrinsic micromotion.
    For this particular measurement,
        the time-dilation shift is $ 8.34 \times \ee$.
    We made several measurements on different days 
        and take the average as our final result.
    }
    %\begin{tabular}
    %\begin{ruledtabular}
    \begin{tabular}{rcccccc}
        \noalign{\smallskip}\hline\noalign{\smallskip}
        Mode &  $\hat{x}$-COM&  $\hat{x}$-STR&  $\hat{y}$-COM&  $\hat{y}$-STR&   $\hat{z}$-COM&  $\hat{z}$-STR\\
        \hline \noalign{\smallskip}
        Frequency(MHz) &         4.00 &      2.84 &      4.03 &      2.84 &      1.17 &      2.10   \\
        $z$(nm) &                 6.6 &       0.6 &       6.4 &       0.5 &       7.1 &       7.8   \\
        $\overline{n}_m$ &        5.2 &       7.0 &       3.0 &       5.7 &      12.8 &       7.5   \\
        $\overline{n}_c$ &        3.0 &       4.6 &       3.0 &       4.6 &       7.3 &       4.0   \\
        TDS/quantum($\ee$) &    0.747 &     0.004 &     0.757 &     0.003 &     0.030 &     0.118   \\
        Total TDS($\ee$) &       3.94 &      0.03 &      2.29 &      0.02 &      0.39 &     0.89    \\
        \noalign{\smallskip}\hline
    \end{tabular}
    %\end{ruledtabular}
\end{table*}

The time dilation shift due to the motion of the $\Al$ ion 
    contributed to most of the uncertainties among all published works 
    on the $\Al$ ion optical clock~\cite{chou_2010_frequency,brewer_2019_clock}.
There are two types of motions for trapped ions:
    micromotion that is driven by the trap RF field and
    harmonic-oscillator (secular) motion at lower frequencies.
In both cases, as the ion moves inside an electric field, 
    the total frequency shift needs to include a frequency-dependent term 
    that corresponds to the Stark effect~\cite{chou_2010_frequency}:
\begin{equation}
    \frac{\Delta v}{v} = - \frac{E_p}{mc^2} \left( 1+\frac{f}{400\textrm{MHz}} \right)^2,
\end{equation}
where $E_p$ and $f$ are the energy and frequency of this motion respectively,
and $m$ is the mass of the $\Al$ ion.

The secular motion energy is dominated by laser cooling.
All the cooling lasers are kept on during the clock interrogation to ensure
    the ions stay close to the sympathetic Doppler cooling limit.
Consider a clock ion with mass $m_2$ = $m$ cooled by another ion with mass $m_1$,
    the secular motion energy of the clock ion can be written as:
\begin{align}
\begin{split}
    E_\textrm{SM,i} 
    & = \zeta_i m \omega_i^2 z_i^2 ( \overline{n}_i + 1/2) \\
    & = \textrm{TDS}_i ( \overline{n}_i + 1/2),
\end{split}
\end{align}
where $\textrm{TDS}_i$ stand for the calculated value representing the
    time dilation shift per quantum number.
$\zeta_i$ is a factor that describes the intrinsic micromotion (IMM)
    driven by the trapping RF field.
$z_i = b_i z_{0,i}$ is the mode amplitude of this motion at the ground state,
    $\omega_i$ is the mode frequency,
    $z_{0,i}=\sqrt{\hbar/(2m_2\omega_i)}$, 
    $b_i$ is the component of the normalized eigenvector for the mode%
    ~\cite{wubbena_2012_sympathetic}, 
    and $\overline{n}_i$ is the average motional quantum number 
    that can be measured by comparing the amplitude of the red and blue sidebands~\cite{cui_2018_sympathetic}:
\begin{equation}
    \overline{n}_i = \frac{P_{r,i}}{P_{b,i}-P_{r,i}}.
\end{equation}

The IMM has a frequency exactly the same as that of the driven field 
    and exists even in an ideal Paul trap.
The energy of the IMM is approximately the same as the secular energy
    in the transverse direction for a single ion.
For two co-trapped ions, $\zeta_i$ can be expressed as~\cite{wubbena_2012_sympathetic}:
\begin{align}
& \zeta_{i, \textrm{COM}} = 1 +
    \frac{2\epsilon^2/\mu}{2\epsilon^2/\mu - 2\alpha - (1-\sqrt{\mu}b_1/b_2)}, \\
& \zeta_{i, \textrm{STR}} = 1 +
    \frac{2\epsilon^2/\mu}{2\epsilon^2/\mu - 2\alpha - (1+\sqrt{\mu}b_2/b_1)}.
\end{align}
COM and STR indicate for center-of-mass mode and stretch mode, respectively.
$\mu = m_2/m_1$,
$\alpha \approx 0.622$ and $\epsilon \approx 2.778$ are geometric parameters for the trapping field. 

\begin{figure}
    \includegraphics[width=0.9\linewidth]{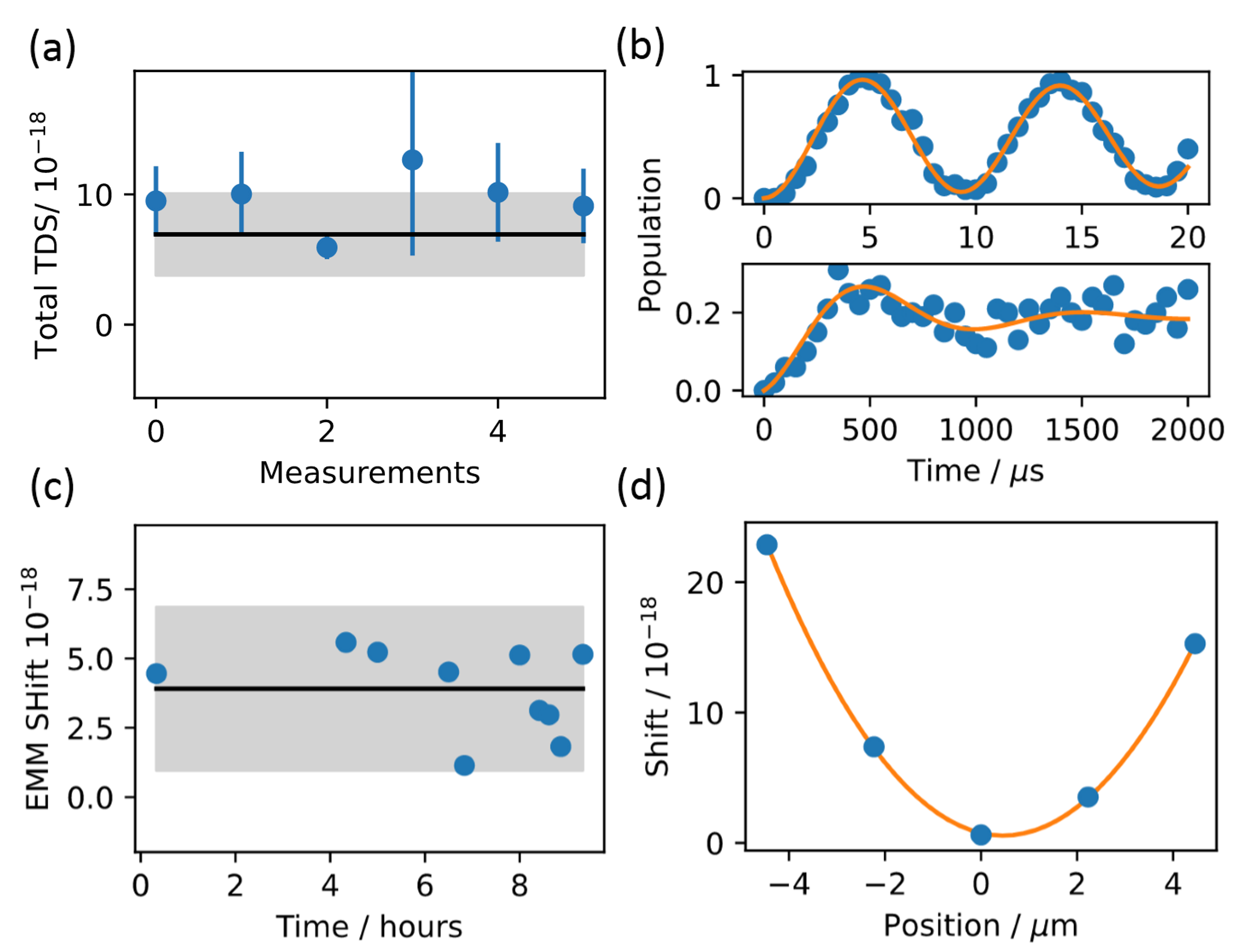}
    \caption{
        Measurements of the time-dilation shift.
        (a) Time-dilation shift due to SM and related IMM measured over several days.
            We took the weighted average as the final result 
            and estimated the uncertainty of this measurement 
            as twice of the standard deviation.
        (b) An example of the Rabi flopping on the carrier (top panel) 
            and the micromotion sideband (bottom panel) 
            on the $\Ca$ ion driven by the 729-vertical beam.
        (c) Average EMM frequency shift measured during a day.
            The clock is free-running without an EMM servo after the initial EMM compensation.
            Moreover, the order of our ion pair is not controlled.
            The final EMM frequency shift is $-(\sEMM \pm \uEMM) \times \ee$,
            where the uncertainty is given 
            by twice of the standard deviation,
            and is shown in the gray band.
        (d) EMM shift along the trap axis.
            As we did not control the order of the $\Ca$ and $\Al$ ions,
            we tried to arrange the ions symmetrically around the minimal micromotion point.
    }
    \label{fig:motion}
\end{figure}

We took several measurements in different days 
    and used the weighted average $-(\sSM \pm \uSM) \times \ee$ as the final result.
The uncertainty is given by twice of the standard deviation of the measurements 
    and is shown in the gray band in Fig. \ref{fig:motion} (a).
One of these measurements is shown in Table \ref{tab:sm_result}.
The secular motion energy at this cooling limit can be calculated~\cite{wubbena_2012_sympathetic}:
\begin{equation}
    E_{i,limit} = \frac{\hbar \Gamma}{24} \frac{2(1+3l_i^2)}{l_i^2},
\end{equation}
where $\Gamma=20.4$ MHz is the natural linewidth of the 
    $\sCa \rightarrow \pCa$ transition on $\Ca$ that is used for the Doppler cooling.
$l_i$ represents the projection of the cooling laser to $i$-th direction.
The calculated cooling limited is listed as $\overline{n}_c$ in Table \ref{tab:sm_result}.
It can be seen that our clock was operated close to this Doppler cooling limit.
This corresponding to a time dilation shift of $-6.3 \times \ee$,
    lower than a similar $\Al$ clock that symmetrically cooled by $\Mg$~\cite{chou_2010_frequency}.
The accuracy of the measurement is limited by the decoherence due to magnetic field noise since it reduces $P_{b,i}$ and $P_{r,i}$.
Cooling the ions much closer to the motional ground states will leads to a higher $P_{b,i}$ and a lower $P_{r,i}$, results in a stronger resistance to the decoherence.
Due to a suitable mass ratio between $\Ca$ and $\Al$ ions, 
we note that two of these motional modes have much smaller amplitude and therefore contribute less to the motional energy of the $\Al$ ion.
This makes it easier to reduce the time-dilation shift due to the secular motion through ground state cooling in the future.

In addition to the IMM, the ion may suffer from additional kinetic energy 
    arising from the imperfection of the trapping potential or phase shifts between the trap electrodes.
This causes the excess micromotion and its energy $E_\textrm{EMM}$ can be measured 
    through the ratio of the Rabi rate of the carrier and the EMM sideband 
    $\eta$~\cite{berkeland_1998_minimization}:
\begin{equation}
    E_{\textrm{EMM}} = \sum_i \frac{m \Omega_{RF}^2 \eta_i^2}{k_L^2 c^2},
\end{equation}    
$i$ is summed over all three perpendicular directions,
    and $k_L$ is the wave vector of the detection laser beam.

Our measurement of $E_\textrm{EMM}$ is performed on the 
    $\Ca$ ion ($m_\textrm{Ca} = m_1$) using three different 729 nm laser beams.
The vertical (729-vertical) and axial (729-axial) beam are perpendicular to each other, 
    while the horizontal beam (729-horizontal) is at 45$^{\circ}$ angles to the axial beam (Fig. \ref{fig:system}).
Taking this into account, the frequency shift on $\Al$ ion can be written as:
\begin{equation}
    E_\textrm{EMM} = 
    \sum_i \frac{m_1^2 \nu_{RF}^2}{m_2 \nu_1^2} \left(\chi_i\eta_i\right)^2,
\end{equation}
where $\nu_1 \approx$ 411.042 THz represents the frequency of 
    the detection laser that excited the $\sCa \rightarrow \dCa$ transition of the $\Ca$ ion.
$\chi_i$ denotes a factor that describes the projection of laser direction.

During the clock operation, 
    we started from minimizing the EMM shift by adjusting the compensation voltages,
    and left it free-running for the rest of the day.
Therefore, we continued measuring the EMM shift throughout the entire day 
    and found the total frequency shift due to EMM 
    by averaging those data, yielding a result of $-(\sEMM \pm \uEMM) \times \ee$.
The uncertainty is given by twice of the standard deviation, 
    as shown in the gray band in Fig. \ref{fig:motion} (c).
    
In a perfect linear Paul trap, EMM does not exist along the trap axis.
However, our trap is not perfect, 
    as we observed a minimum axial EMM at a special location (Fig.~ \ref{fig:motion} (d)).
It's hard to keep the $\Al$ ion staying at this point because random background-gas collision
    switched the order of the $\Ca-\Al$ pair approximately every 1000 s.
We did not control the order of the ions,
    instead, we moved the center of the ion pair close to this minimum point 
    and left it free running.
We observed some reorder events during the clock operation, and the EMM measurements were taken in both orders randomly.
Other sources that lead to a change of the EMM, such as ion reloading or charging due to the clock laser\cite{brewer_2019_clock} are not evident in our case since the change of the $\Al$ ion's position due to reordering event is the dominant effect.

% Background gas collisions shift
The reorder events are recorded and used to evaluate the collision shift for the background gas.
The energy that is required by the $\Ca-\Al$ ions to change their order is given by~\cite{hankin_2019_systematic}:
\begin{multline}
    E_{reorder} = \frac{3}{4} \left( \frac{\sqrt{m}\omega_z e^2}{2\pi\epsilon_0}\right)^{2/3} \\
    \times 
    \left( \left\{ 
        \frac{2(\epsilon^2+\alpha-1)[\epsilon^2+\mu(\alpha-1)]}{
        \epsilon^2(\mu+1)+2\mu(\alpha-1)} 
    \right\}^{1/3} -1 \right),
\end{multline}
where $e$ is the charge of the electron. 
$\alpha, \epsilon$ are geometric factor that associate with the trap itself.
$\mu=m_2/m_1$ is the ratio of $\Al$ and $\Ca$ ion's mass.
The relationship between the vacuum pressure and the reorder rate is given by:
\begin{equation}
    \Gamma_{reorder} \lesssim \frac{1}{2} 
        \left(\frac{p}{902 \textrm{nPa}}\right)
        \left( \frac{E_{reorder}}{1\textrm{K}\times \textrm{k}_\textrm{B}} \right).
\end{equation}

The reorder event can be easily observed 
    on the EMCCD camera  
    since the position of the $\Ca$ ion is shown as a bright while the $\Al$ ion is invisible. 
We took an image from the EMCCD every 0.5 s and averaged 10 images to reduce noise from the background.
This averaging normally does not miss any reorder events, 
    as the reorder period is around 1000 s, much longer than the imaging time.
The geometric center of the measured $\Ca$ ion was recorded to determine the position of the ion.

The reorder rate is given by $\Gamma_{reorder} = N_{reorder}/T$, 
    where $N_{reorder}$ is the number of reorder events during time $T$. 
In our case, the reorder rate was measured to be $\Gamma_{reorder}=$0.0013 s${}^{-1}$, 
    corresponding to a vacuum pressure of 3.8 nPa. 
Based on an empirical formula from Rosenband et al.~\cite{rosenband_2008_frequency}, 
    and assuming that each collision brings a maximum phase shift of $\pi/2$, 
    background gas collision leading a frequency shift up to 0.195 mHz,
    corresponding to a fractional frequency shift of $\uBGC \times \ee$.
    We take this value as the uncertainty of this effect.

\begin{table*}
    \caption{\label{tab:temperature}
    The result of the systemic shift due to black-body radiation (BBR) of the ion trap system.
    }
    %\begin{ruledtabular}
    \begin{tabular}{rcccc}
        \hline
        Components 
        &  $\Omega_{eff}$ 
        &  Value (K) 
        &  Uncertainty (K) 
        &  Uncertainty $T_{eff}$ (K) \\
        \hline
        Blade electrodes         & 0.457     & 303.87    & 2.90      & 0.110     \\
        Cap electrodes           & 0.293     & 304.16    & 2.59      & 0.064     \\
        Insulation supports      & 2.284     & 302.78    & 2.58      & 0.488     \\
        Compensation electrodes  & 0.245     & 304.07    & 2.65      & 0.055     \\
        Stainless steel brackets & 0.675     & 302.78    & 2.21      & 0.128     \\
        Chamber                  & 1.264     & 297.97    & 1.43      & 0.169     \\
        Glass Windows            & 6.016     & 297.97    & 1.43      & 0.801     \\
        Flanges                  & 1.333     & 297.97    & 1.43      & 0.178     \\ 
        \hline
    \end{tabular}
    %\end{ruledtabular}
\end{table*}    

The blackbody radiation (BBR) leads to an AC Stark shift in the clock transition.
The $\Al$ clock is operated at room temperature ($\approx$~300~K), 
The BBR temperature is mainly caused by the thermal radiation emitted 
    from various components of the ion trap. 
We built another ion trap system with the same configuration 
    to achieve an accurate assessment of the BBR temperature 
    using an infrared thermal imager.
Temperature sensors were placed on both this trap and the actual system to ensure that they work in the same condition.
The finite element analysis method was also employed 
    to achieve an accurate assessment of the BBR temperature~\cite{zhang_2021_evaluation}.
For the $\Al$ ion, the effective BBR temperature can be calculated using:
\begin{equation}
    T_{eff} = \sqrt[4]{\sum_i \frac{\Omega_{eff}}{4\pi}T_i^4},
\end{equation}
where $\Omega_{eff} $ and $ T_i $ are the effective solid angle 
    and temperature of each component of the trapped ion system listed in Table~\ref{tab:temperature}.
This gives $T_{eff}$ = 299.6 (1.0) K.
We noticed that there is a temperature difference of $0.94$K between the simulation trap and the actual trap when driving with the same RF field.
Taking that into consideration, the effective BBR temperature felt by the ion is estimated to be $T_{eff} = 299.6 \pm 1.4$ K. 
The clock frequency shift due to BBR is calculated using~\cite{brewer_2019_clock}:
\begin{equation}
    \Delta \nu_{BBR} = 
    -\frac{\pi \Delta \alpha_0}{60 \epsilon_0 c^3} \left( \frac{k_B^4 T_{eff}^4}{\hbar^4} \right)^4,
\end{equation}
where $\Delta\alpha_0 = 7.03(94) \times 10^{-42}$ Jm$^2$/V$^2$ 
    is the static differential polarizability. 
The corresponding BBR shift is  $-(\sBBR \pm \uBBR) \times \ee$.

The electric field of the optical radiation incident on the ion 
    perturbs and shifts the line center of the ion transition. 
This contributes to the Stark shift~\cite{brewer_2019_clock}:
\begin{equation}
    {\Delta v}_{ac} = -\frac{\Delta\alpha_{ac}(\lambda)}{2 h c \epsilon_0} I, 
\end{equation}
where $ \Delta\alpha_{AC}(\lambda) $ is the dynamic polarizability of the wavelength 
    $ \lambda $ and I is the intensity of the incident light. 
The intensity of the clock laser was measured to be 40 nW, focused on an area  
    of 120 \textmu m in diameter, 
    corresponding to an AC Stark shift $(\sClock \pm \uClock) \times \ee$.

In addition to the clock laser itself,
    both the 397 nm cooling light 
    and the 866 nm pumping light are present during clock operation.
The intensities of both beams are calibrated by monitoring the fluorescence of the ions periodically during the clock operation.
We noticed that the fluorescence will change significantly if the power of the 397 nm laser exceeds 34 nW.
With a measured beam wrist of 80 \textmu m, this lead to an AC stark shift up to $-1.3 \times 10^{-19}$.
Since the 866 nm beam and the 729-horizontal beam share the same fiber and optics, the waist position of these beams are different due to the optical dispersion.
To ensure the 729-horizontal beam has the strongest coupling with the ion, the ion have to stay approximately 100 \textmu m off from the 866 nm beam waist.
This leads to a much higher requirement of the 866 nm laser power, reaching 7.4 \textmu W.
We checked the possible maximum laser intensity around the ion (within a range of 20 \textmu m $\times$ 20 \textmu m) using beam profiler and figured out the maximum possible AC stark shift for the 866 nm laser can reach $-9.4 \times 10^{-19}$.
In summary, the total AC Stark shift due to these three lasers is smaller than $1.1\times 10^{-18}$.

%\subsection{AOM phase chirp}
Phase chirp in the clock beam AOM can also contribute to a frequency shift
    as the optical path through the crystal changes when it switches on and off.
The shift due to a phase chirp is
minimized by using a very lower RF driving power (1.1 mW) on the AOM~\cite{rosenband_2008_frequency}.
This lead to a shift smaller than $\uAOM \times \ee$.

% 1-st Dopper
\begin{figure}
    \includegraphics[width=0.95\linewidth]{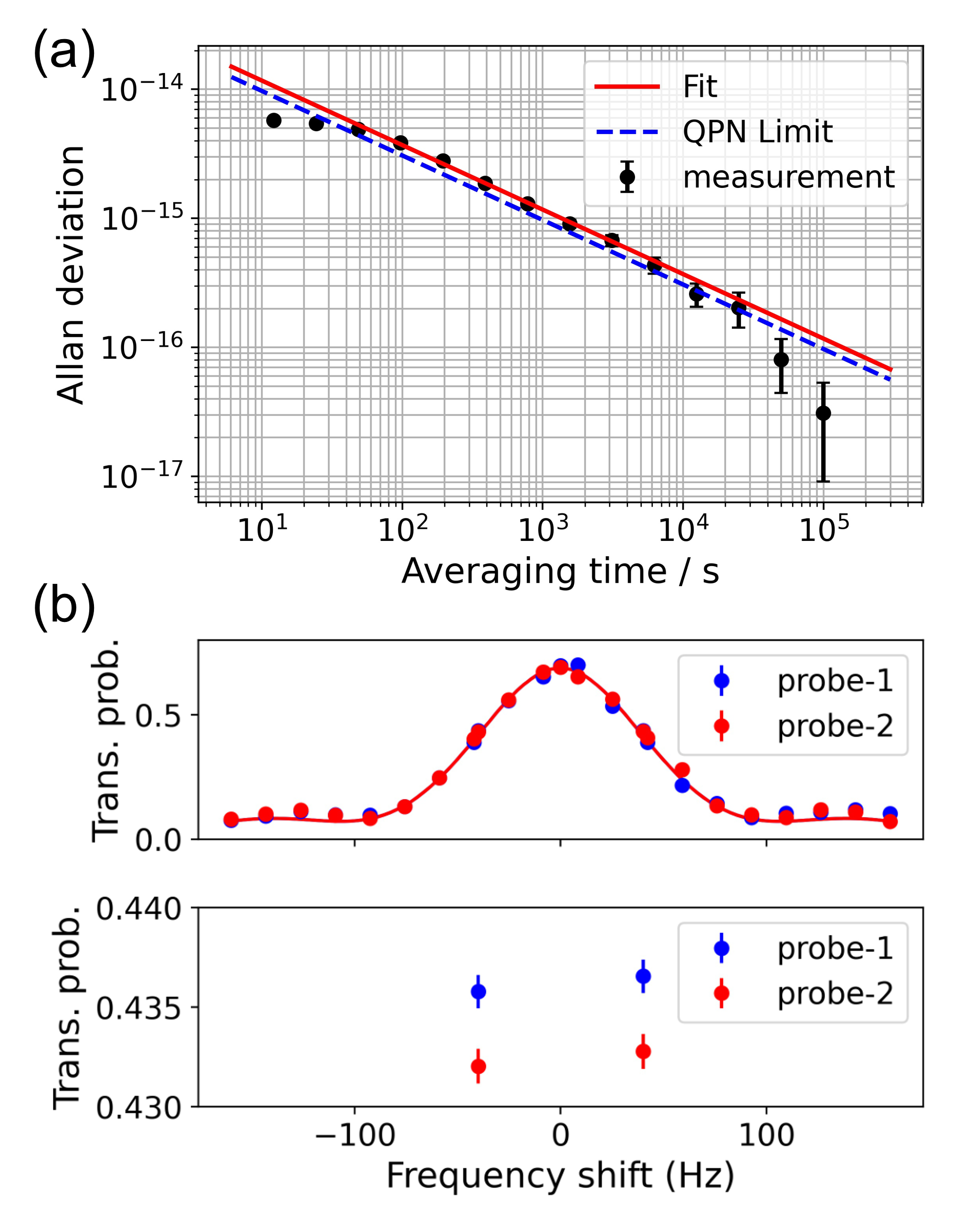}
    \caption{
        (a) The stability of the clock determined from comparing the output frequency of laser beam probe-1 and probe-2.
        The data was taken after an operation time of \~160 hours, with a duty cycle of 48\%.
        Red line is a fit using $\sigma_y/\sqrt{\tau}$ while $\sigma_y=3.7\times10^{-14}$.
        The quantum projection noise (QPN) is shown in blue dashed trace, $\sigma_{QPN}=3.2\times10^{-14}$, has taken account into the effect from the loss of contrast due to the decoherence from the vibrational noise of the laser path.
        (b) Interleave detection is employed during the comparison to visualize the locked line-shape from both laser beam directions. The lower panel is a zoom in of the transition probability of the left and right side of the peak.
        The unbalanced probabilities suggested a residual first order Doppler shift from the drifting laser, which is smaller than $0.6\times10^{-19}$. 
        }
    \label{fig:1st_doppler}
\end{figure}

% 1st doppler shift
The first-order Doppler shift cannot be observed 
    for most optical clocks that work in the Lamb-Dicke regime 
    because the motional amplitude is much smaller than the laser wavelength. 
However, a first-order Doppler shift may
    still occur when the ion itself moves in a fashion correlated 
    with the clock laser. 
This movement may come from 
    the contraction of the structure of the ion trap 
    or an additional electric field generated by the UV photoelectric effect.
We compared the output frequency of the counter-propagating probe-1 and probe-2, looking for the first order Doppler shift. 
No statistical frequency difference can be observed with a fractional uncertainty of $5.4\times 10^{-17}$.
Interleaved detection is employed during the comparison to show the Rabi line-shape seen from these two directions (Fig.~3 (b)).
When the average frequencies of these two directions are used as the output of the clock, possible first order Doppler shift can be further canceled.
However, due to the drift of the laser locking on the cavity, and the imperfectly matched laser intensity, a residual frequency shift may arise from the imbalanced transition probabilities of the left and right sides of the peak~\cite{rosenband_2008_frequency}.
This lead to a frequency shift smaller than $\uDoppler \times 10^{-19}$, limited by the servo error of these two feedback loops.

The instability of the comparison is calculated from the measured frequency difference, shown in black dots in Fig.3 (a).
Fitting of these data points gives $\sigma_y(\tau) = 3.7\times10^{-14}/\sqrt{\tau}$ and marked as a red line.
The experiment lasted for \~160 hours, with a duty cycle of 48\%.
Although the $\Ca$-$\Al$ ion pair can remind in the trap for several days without running the experiment,
probing the $\s \leftrightarrow \p$ transition will shorten the lifetime of the $\Al$ ion to approximately 2 hours.
This is because the $\Al$ ion has a much higher possibility of reacting with background hydrogen molecules when excited in the $\p$ state.
Reloading and calibrating process cost \~20 minutes when an aluminum hydride molecule is formed.
However, laser drifts, especially the drift from the cooling lasers require manual adjustments during the lock, which now limits the improvements of the duty cycle time.
    
In summary, we have evaluated the fractional frequency shift 
    and associated systematic uncertainty 
    for an $\Al$ clock sympathetically cooled by a $\Ca$ ion. 
The total shift is 
    $-(\sTotal \pm \uTotal) \times \ee$.
The systematic uncertainty is limited by the quadratic Zeeman shift, 
    which is mainly caused by the measurement uncertainty 
    of the quadratic Zeeman coefficient $C_2$. 
Significant improvements can be made when the magnetic field $\braket{B}$ is reduced in the future.
Our measurement of the secular motion temperature is also limited by decoherence
    from magnetic field noise and jitter on the amplitude of the RF trapping field.
In addition, controlling the order of the ions is inevitable 
    that the uncertainty of time dilation shift 
    due to excess micromotion can be improved.
Owing to the relatively lower Doppler cooling limit from the $\Ca$ ion
    and the suitability mass ratio to $\Al$ ion, 
    we achieved a lower time-dilation shift due to secular motion 
    than that of a $\Mg - \Al$ clock. 
As the $\Ca$ ion has a suitable energy level for electromagnetically-induced-transparency (EIT) cooling,
it is possible to bring the motional energy much closer to the motional ground state to reduce the time-dilation shift due to the motion.
Benefits from its simple laser system~\cite{cao_2022}, optical clocks based on $\Ca$ - $\Al$ ions have the potential to achieve a transportable, compact design.

\section*{Author contribution}
S.W, J.C, H.S and X.H developed components of the experimental apparatus.
P.Z, Y.W evaluated the black-body radiation shifts.
K.C, S.C, C.S, Y.W and J.Y collected and analyzed most of the data. 
All authors discussed the results and contributed to the writing of the paper.

\section*{Data availability} The datasets generated during and/or analysed during the current study are available from the corresponding author on reasonable request.

\begin{acknowledgements}
We thank M. Zhan and C. Li for useful discussions.
This work is supported by 
    the National Key R\&D Program of China (Grant No. 2017YFA0304401),
    the Strategic Priority Research Program of the Chinese Academy of Sciences (Grant No. XDB21030100),
    the Technical Innovation Program of Hubei Province (Grant No. 2018AAA045)
    and the National Natural Science Foundation of China (Grant No. 11904387, U21A20431).
\end{acknowledgements}

\bibliographystyle{spphys}       % APS-like style for physics
\bibliography{evaluation}   % name your BibTeX data base

\end{document}